\documentclass[%
 preprint,
 amsmath,amssymb,
 aps,
]{revtex4-1}

\usepackage{graphicx}
\usepackage{dcolumn}
\usepackage{bm}
\usepackage[flushleft]{threeparttable}
\usepackage{url}
\usepackage{amsmath}
\usepackage{amssymb}
\usepackage{braket}
\usepackage{multirow}
\usepackage{array}
\usepackage{url}
\usepackage{booktabs}
\newcolumntype{P}[1]{>{\centering\arraybackslash}p{#1}}

\begin{document}

\title{Low Operation Voltage Ferroelectric Field-Effect Transistor Based on Polarization Rotation Effect}

\author{Y. Qi and A. M. Rappe}
\affiliation{%
 The Makineni Theoretical Laboratories, Department of Chemistry,\\
 University of Pennsylvania, Philadelphia, PA 19104-6323 USA\\
}%
\date{\today}

\begin{abstract}
The effect of polarization rotation on the performance of metal oxide semiconductor field-effect transistors was investigated with a Landau-Ginzburg-Devonshire theory based model.
In this analytical model, depolarization field, polarization rotations and the electrostatic properties of the doped silicon substrate are considered 
to illustrate the size effect of ferroelectric oxides and the stability of polarization in each direction. 
Based on this model, we demonstrate that MOSFET operation could be achieved by polarization reorientation with a low operating voltage,
if the thickness of ferroelectric oxide is properly selected.
Polarization reorientation can boost the surface potential of the silicon substrate, leading to a subthreshold swing $S$ lower than 60 mV/decade.
We believe that this model could provide guidance in designing electronic logic devices with low operating voltages and low active energy consumption.
\end{abstract}

\pacs{Valid PACS appear here}
\maketitle
\section{Introduction}

Ferroelectric oxides are a promising class of materials for application in electronic devices due to their intrinsic spontaneous electric polarization, 
which can not only control the conductance of the channel, but also can be reoriented by an external electric field~\cite{Mathews97p238,Naber05p243,Hoffman10p2957,Fu00p281}. 
By modulating the polarization of ferroelectric oxides, 
programmable binary logic devices can be achieved, and the fast reorientation of polarization enables fast switching and lower--power operation of the
metal--oxide--semiconductor field--effect transistor (MOSFET)~\cite{Kimura04p919,Liu13p053505}.

Here, we aim to provide guidance about designing a programmable fast switching MOSFET with proper size,
by considering factors which were rarely included in previous modeling, 
but may strongly affect the polarization reorientation and size effect of ferroelectric oxides.
Many analytical models based on the Landau-Ginzburg-Devonshire (LGD) theory have been proposed previously~\cite{Chen12p110,Chen11p2401} to simulate the electrical behaviors of MOSFETs. 
These models provide an insightful understanding about the mechanisms of ferroelectric oxide based MOSFETs and guide the fabrication of novel devices. 
However, there are still several vital factors beyond the scope of previous models.
First, the effect of the polarization distribution in three dimensions (3D) on the channel current--gate voltage relationship of a MOSFET was rarely considered, 
even though there were a lot of studies about polarization in 3D and its response to electric fields in different orientations~\cite{Bellaiche01p060103,Hlinka06p104104,Fu00p281,Nagarajan02p43}. 
For simplicity, the polarization of the ferroelectric oxide in a MOSFET is usually treated in one dimension.    
It is true that the channel conductance is mainly modulated by the out--of--plane polarization, 
but it is also important to note that the polarization components in all three dimensions are coupled together, and the in--plane polarization strongly influences the electric susceptibility out of plane. 
The second factor is the electrostatic properties of the channel and gate electrode. It is widely known that the distribution of charge in electrodes, 
which is parameterized as screening length,
determines the strength of depolarization field, which affects the magnitude of the spontaneous ferroelectric polarization~\cite{Batra73p3257,Sai05p020101R,Al-Saidi10p155304,Mendez-Polanco12p214107}.

In this letter, we propose a LGD theory based single crystal model with detailed analysis of these factors,
in order to provide strategies for designing a low operation voltage ferroelectric field-effect transistor~\cite{Datta13p6,Salahuddin08p405}.
A previous study argues that a subthreshold swing lower than 60 mV/decade can be achieved by the negative capacitance effect~\cite{Salahuddin08p405}, 
but there is also debate that direct current negative capacitance is not possible, due to Gibbs free energy considerations~\cite{Krowne11p988}.
In our model, the polarization dynamics obeys the Landau--Khalatnikov equation and is always minimizing the Gibbs free energy under a unidirectional gate voltage.
We demonstrate that fast switching (subthreshold swing lower than 60 mV/decade) can be achieved by a proper design of the ferroelectric oxide size. 
The mechanism is that during the process of polarization reorientation, 
the tendency to possess spontaneous polarization in a new direction boosts the screening charge accumulation and channel current increases, 
leading to a low subthreshold swing.

\section{Model Approach}
The LGD model is a phenomenological theory which describes the electrical properties of ferroelectric oxides. 
In this model, the thermodynamic potential (Gibbs free energy $G_{0}$) of a single crystal ferroelectric oxide is given as a function of polarization in three directions~\cite{Pertsev98p1988,Koukhar01p214103}.
\begin{equation}
\begin{aligned}
&G_{0}=\alpha_{1}\left(P_{x}^{2}+P_{y}^{2}+P_{z}^{2}\right)+\alpha_{11}\left(P_{x}^{4}+P_{y}^{4}+P_{z}^{4}\right)\\
&+\alpha_{12}\left(P_{x}^{2}P_{y}^{2}+P_{y}^{2}P_{z}^{2}+P_{z}^{2}P_{x}^{2}\right)+\alpha_{111}\left(P_{x}^{6}+P_{y}^{6}+P_{z}^{6}\right)\\
&+\alpha_{112}\left[P_{x}^{4}\left(P_{y}^{2}+P_{z}^{2}\right)+P_{y}^{4}\left(P_{z}^{2}+P_{x}^{2}\right)+P_{z}^{4}\left(P_{x}^{2}+P_{y}^{2}\right)\right] \\
&+\alpha_{123}P_{x}^{2}P_{y}^{2}P_{z}^{2} \\
\end{aligned}
\end{equation}
Taking external electric field and internal depolarization field into consideration, electrostatic terms should be added as
\begin{equation}
G=G_{0}-E_{x}P_{x}-E_{y}P_{y}-E_{z}P_{z}
\end{equation}

FIG. 1. shows the schematic of the MOSFET that we study. 
The insulator between the gate electrode and doped silicon substrate is a ferroelectric oxide~\cite{Salahuddin08p405,Ngo14p082910}.
The $z$ axis is normal to the ferroelectric oxide/silicon interface.
In the $x$ and $y$ directions, there is no external voltage, and short circuit conditions are applied~\cite{Batra73p3257,Mehta73p3379,Kolpak06p054112}.
\begin{figure}[htbp]
\centering\includegraphics[width=8.0cm]{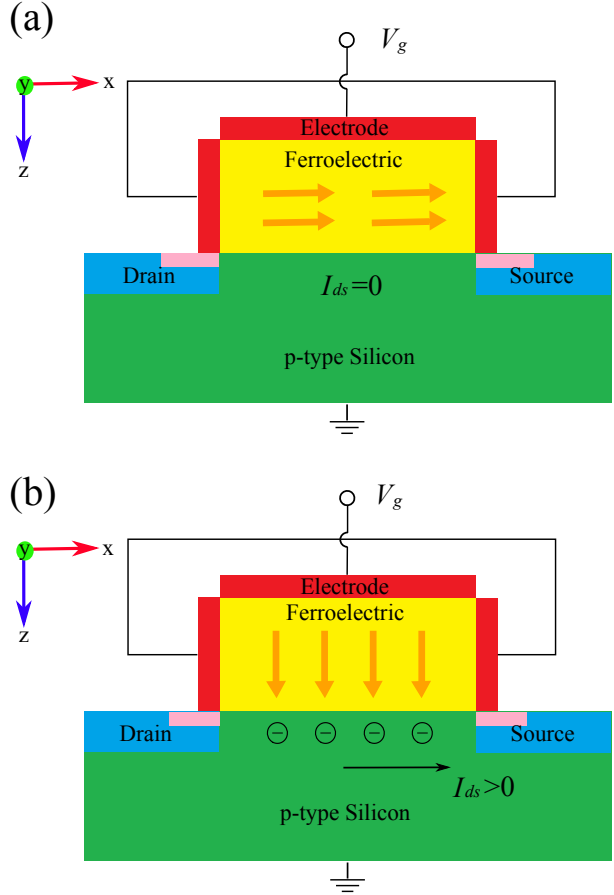} 
\caption{The schematic of a MOSFET with ferroelectric oxide as the insulator between the gate electrode and silicon substrate. 
The pink rectangles represent insulator layers isolating the side electrodes and source, drain terminals. 
(a) No gate voltage is applied $\left(V_g=0\right)$ and the polarization is in plane. No carriers or current is in the channel; 
(b) Gate voltage is applied $\left(V_g>0\right)$ and polarization is out of plane. Carriers are induced by the polarization, and drain--source current flows.}\label{fig:1}
\end{figure}
For the case that a gate voltage $V_{g}$ is imposed on the MOSFET, we have the following equations:
\begin{equation}
\left.
\begin{aligned}
&2V_{ex}+V_{ox}=0  \\
&2V_{ey}+V_{oy}=0  \\
&V_{ez}+V_{oz}+\varphi_{s}=V_g  \\
\end{aligned}
\right\}
\end{equation}
$V_{ex,ey,ez}$ and $V_{ox,oy,oz}$ are the voltage drop across the electrode and the ferroelectric oxide in the $x$, $y$ and $z$ directions.
$\varphi_{s}$ is the surface potential of the silicon substrate, and it can also be viewed as the voltage drop in the doped silicon substrate.
The flat band potential $V_{fb}$, which results from the alignment of the Fermi levels of the gate electrode, oxide, and silicon substrate is included in $V_g$.
The electric field $\bm{E}$ is determined by both the external applied voltage and the electrostatic properties of the ferroelectric oxide and electrodes~\cite{Wurfel73p5126,Junquera03p506,Kim05p237602}.
It is widely accepted that the charge density in noble metal electrodes follows the Thomas-Fermi distribution, 
and this distribution causes a voltage drop across the electrodes.
The following derivation calculating this potential drop follows the main idea in Ref.~\cite{Mehta73p3379}, but is re-interpreted.
Taking the $z$ direction as an example and the electrostatic properties in the $x$, $y$ directions following similar rules, the relationship between the electric field and the charge density takes the form
\begin{equation}
\frac{dE\left(z\right)}{dz}=-Q\left(z\right)=-q\frac{n\left(z\right)-n_0}{\varepsilon_0\varepsilon_e}
\end{equation}
$E\left(z\right)$, $Q\left(z\right)$ and $n\left(z\right)$ are the electric field, the charge density and the electron density in electrodes at the position $z$. 
$n_0$ is the average electron density in a neutral electrode. 
$q$ is the electronic charge.
$\varepsilon_0$ and $\varepsilon_e$ are the electric permittivities of the vacuum and electrode. 
Meanwhile, the potential drop $V\left(z\right)$ is expressed as
\begin{equation}
\frac{dV\left(z\right)}{dz}=-E\left(z\right) \quad  \Longrightarrow \quad \frac{dV\left(z\right)}{dn\left(z\right)}\frac{dn\left(z\right)}{dz}=-E\left(z\right) 
\end{equation}
The electrons in metal electrodes are treated as a free Fermi gas, so the local potential and the electron density are related as
\begin{equation}
V=\frac{\hbar^2}{2m}\left(3\pi^2n\right)^{\frac{2}{3}} 
\end{equation}
\begin{equation}
\frac{dV\left(z\right)}{dn\left(z\right)}=\frac{\hbar^2}{3m}\left(3\pi^2\right)^{\frac{2}{3}}n\left(z\right)^{-\frac{1}{3}} 
\end{equation}
$\hbar$ is the reduced Planck constant and $m$ is the electronic mass.
By combining equations (5) and (6), we have
\begin{equation}
\frac{dn\left(z\right)}{dz}=-\left[\frac{\hbar^2}{3m}\left(3\pi^2\right)^{\frac{2}{3}}n\left(z\right)^{-\frac{1}{3}}\right]^{-1}E\left(z\right)
\end{equation}
Taking the derivative of equation (4), we have
\begin{equation}
\frac{d^2E\left(z\right)}{dz^2}=-\frac{q}{\varepsilon_0\varepsilon_e}\frac{dn\left(z\right)}{dz}=\frac{3mq}{\varepsilon_0\varepsilon_e{\hbar^2}}\left(3\pi^2\right)^{-\frac{2}{3}}n\left(z\right)^{\frac{1}{3}}E\left(z\right)
\end{equation}
The characteristic length ${\lambda}_{z}$, which is also called screening length and determines the dispersion of electrons in electrodes, is defined as
\begin{equation}
{\lambda}_{z}^2=\left[\frac{3mq}{\varepsilon_0\varepsilon_e{\hbar^2}}\left(3\pi^2\right)^{-\frac{2}{3}}n\left(z\right)^{\frac{1}{3}}\right]^{-1}
\end{equation}
Therefore, equation (9) is rewritten as
\begin{equation}
\frac{d^2E\left(z\right)}{dz^2}=\frac{1}{{\lambda}_{z}^2}E\left(z\right)
\end{equation}
The boundary conditions are
\begin{equation}
\left\{
\begin{aligned}
&E\left(0\right)=\frac{Q_{z}}{\varepsilon_0\varepsilon_e}     \\    
\\
&E\left(-\infty\right)=0     \\
\end{aligned}
\right.
\end{equation}
$Q_{z}$ is the screening charge density at the ferroelectric/oxide electrode interface, which is also $Q\left(z=0\right)$. Thus, the electric field and potential drop through one electrode are
\begin{equation}
E\left(z\right)=\frac{Q_z}{\varepsilon_0\varepsilon_e}e^{z/{\lambda}_{z}}
\end{equation}
\begin{equation}
V_{e}=\int_{-\infty}^0E\left(z\right)dz   
=\int_{-\infty}^0\frac{Q_{z}}{\varepsilon_0\varepsilon_e}e^{z/{\lambda}_{z}}dz=\frac{Q_z{\lambda}_{z}}{\varepsilon_0\varepsilon_e}
\end{equation}
The heterostructure of electrode, ferroelectric oxide, and silicon substrate can be regarded as a capacitor, with equal charge densities at each interface. 
However, the charge distribution in the doped silicon substrate is quite different from that in metal. This is because electrons in the metal are treated as a free electron gas.
This is the basic approximation of the Thomas-Fermi model. 
But doped silicon is a semiconductor, and the free carrier density is local potential dependent~\cite{sze2006physics,taur1998fundamentals}.
The interface charge density-potential relationship in the silicon substrate is given by
\begin{equation}
Q_z=\sqrt{2\varepsilon_{Si}kTN_a}\cdot
\left[\left(e^{-\frac{q\varphi_s}{kT}}+\frac{q\varphi_s}{kT}-1\right)+\frac{n_i^2}{N_a^2}\left(e^{\frac{q\varphi_s}{kT}}-\frac{q\varphi_s}{kT}-1\right)\right]^{\frac{1}{2}}
\end{equation}
\begin{equation}
\frac{d\varphi\left(z\right)}{dz}=E\left(z\right)=\sqrt{\frac{2kTN_a}{\varepsilon_{Si}}}
\cdot\left[\left(e^{-\frac{q\varphi\left(z\right)}{kT}}+\frac{q\varphi\left(z\right)}{kT}-1\right)+\frac{n_i^2}{N_a^2}\left(e^{\frac{q\varphi\left(z\right)}{kT}}-\frac{q\varphi\left(z\right)}{kT}-1\right)\right]^{\frac{1}{2}}
\end{equation}
$\varphi_s$ is the surface potential of the silicon substrate. 
$k$ is the Boltzmann constant. Other parameters are listed and described in TABLE $\rm\uppercase\expandafter{\romannumeral1}$.

From the analysis above, we see that the charge density decreases gradually away from the oxide in both the metal electrode and the doped silicon substrate, 
even though the analytical expressions and physical mechanisms which govern the charge distribution are different.
As a result, there are voltage drops through each layer. 
These voltage drops could counteract or completely neutralize the applied gate voltage, exerting significant influence on the magnitudes of ferroelectric polarization and charge in the channel.
Equation $\left(15\right)$ demonstrates that there is a one-to-one correlation between the interface charge density $Q_z$ and the surface potential $\varphi_s$. 
$\varphi_s$ is a function of $Q_z$:
\begin{equation}
\varphi_s=f\left(Q_{z}\right)
\end{equation}
The voltage drop across the ferroelectric oxide takes the form
\begin{equation}
V_{oz}=E_{z}\cdot{d_{z}}=\frac{Q_{z}-P_{z}}{\varepsilon_{0}}d_{z}
\end{equation}
$d_{z}$ is the thickness of the ferroelectric film and $P_{z}$ is the polarization in the $z$ direction. With the analysis above, equation set $\left(3\right)$ is rewritten as
\begin{equation}
\left.
\begin{aligned}
&\frac{Q_x{\lambda}_{x}}{\varepsilon_0\varepsilon_e}+\frac{Q_{x}-P_{x}}{\varepsilon_{0}}d_{x}+\frac{Q_x{\lambda}_{x}}{\varepsilon_0\varepsilon_e}=0  \\
&\frac{Q_y{\lambda}_{y}}{\varepsilon_0\varepsilon_e}+\frac{Q_{y}-P_{y}}{\varepsilon_{0}}d_{y}+\frac{Q_y{\lambda}_{y}}{\varepsilon_0\varepsilon_e}=0 \\
&\frac{Q_z{\lambda}_{z}}{\varepsilon_0\varepsilon_e}+\frac{Q_{z}-P_{z}}{\varepsilon_{0}}d_{z}+f\left(Q_{z}\right)=V_g  \\
\end{aligned}
\right\}
\end{equation}

For short-circuit conditions, in order to balance the potential drop in the electrodes, the sign of $V_{ox,oy}$ should be opposite to that of $V_{ex,ey}$. 
This indicates that surface charge density should be smaller than polarization, which means an incomplete screening of the polarization charge. 
As a result, an electric field (depolarization field) is induced opposite to the polarization. 
The potential drop in the metal electrodes, which is proportional to screening length,
is the origin of the incomplete polarization charge screening and the depolarization field which suppresses ferroelectricity. 

The energy surface versus polarization direction and magnitude can be plotted under the electrostatic restrictions expressed in equation $\left(19\right)$. 
After acquiring the energy surface, polarization dynamics on the energy surface is simulated by the Landau--Khalatnikov equation~\cite{Landau54p469,Lo03p3353,Zhang05p185},
\begin{equation}\label{e1}
\gamma\frac{d\overrightarrow{P}}{dt}+\nabla_{\overrightarrow{P}}G=0
\end{equation}
$\gamma$ is the polarization dynamic parameter. $G$ is the thermodynamic potential defined in equation (2) with the restriction shown in equation (19).
The most stable polarization is the one which minimizes Gibbs free energy $G$. 
However, if the polarization is not in a local minimum, it cannot move to one instantaneously.
The rate of return to a minimum is determined by many factors.
For example, the resistance of the circuits affects this rate, because polarization evolution must be accompanied by screening charge transmission.
The polarization dynamic parameter $\gamma$ is related to the mobility of polarization, 
as $\nabla_{\overrightarrow{P}}G$ can be regarded as the driving force of polarization and $\gamma\frac{d\overrightarrow{P}}{dt}$ is the speed of polarization evolution.
The applied time--varying gate voltage takes the form,
\begin{equation}
V_g={V_0}\sin\left({\omega}t\right)\qquad \left(0<t<\frac{\pi}{\omega}\right)
\end{equation}
Here, we do not mean that the applied gate voltage is oscillatory. Instead, we are simulating one on/off programmable cycle $\left(0<t<\frac{\pi}{\omega}\right)$,
and the increase/decrease of the gate voltage takes the sine form. 
Equation (\ref{e1}) is rewritten as
\begin{equation}
\gamma_0\frac{d\overrightarrow{P}}{d\left(\omega{t}\right)}+\nabla_{\overrightarrow{P}}G=0
\end{equation}
$\gamma_0=\omega\gamma$ is the effective polarization dynamic parameter.
$\varphi_s$ and $Q_z$ can be calculated from $P_z$ and 
the drain-source current $I_{ds}$ is obtained by the Pao-Sah double integral ~\cite{Pao66p927}.
\begin{equation}
I_{ds}=q\mu_{\rm{eff}}\frac{W}{L}\int_0^{V_{ds}}\left(\int_\delta^{\varphi_s}\frac{\frac{n_i^2}{N_a}e^{q\left(\varphi-V\right)/kT}}{\xi\left(\varphi,V\right)}d{\varphi}\right)dV
\end{equation}
where the parameters in this simulation are listed in TABLE $\rm\uppercase\expandafter{\romannumeral1}$, and the function $\xi\left(\varphi,V\right)$ is given in Ref.~\cite{taur1998fundamentals}. 
\begin{table}[htbp]
\caption{PARAMTERS INVOLVED IN THIS STUDY}
\begin{center}
\begin{tabular}{@{}lll@{}} 
\hline
\hline
 & Description & Value \\ 
 \hline
 $T$                 & Temperature                                             &   298 K \\
 $\alpha_1$          & Coefficient in LGD theory$^a$                           &   $-2.77{\times}10^{7}$ m/F  \\
 $\alpha_{11}$       & Coefficient in LGD theory$^a$                           &   $-5.35{\times}10^{8}$ m$^{5}$/C$^{2}$F  \\
 $\alpha_{12}$       & Coefficient in LGD theory$^a$                           &   $3.23{\times}10^{8}$ m$^{5}$/C$^{2}$F \\
 $\alpha_{111}$      & Coefficient in LGD theory$^a$                           &   $8.00{\times}10^{9}$ m$^{9}$/C$^{4}$F  \\
 $\alpha_{112}$      & Coefficient in LGD theory$^a$                           &   $4.47{\times}10^{9}$ m$^{9}$/C$^{4}$F  \\
 $\alpha_{123}$      & Coefficient in LGD theory$^a$                           &   $4.91{\times}10^{9}$ m$^{9}$/C$^{4}$F  \\
 $\lambda_{x,y,z}$   & Screening lengths in noble metal$^b$                    &   0.04 nm \\
 $\varepsilon$       & Dielectric constant of noble metal$^b$                  &   2.0     \\   
 $N_a$               & Substrate doping concentration$^c$                      &   $4\times10^{15}$ cm$^{-3}$ \\ 
 $n_i$               & Intrinsic carrier concentration$^c$                     &   $1.5\times10^{11}$ cm$^{-3}$ \\ 
 $\varepsilon_{Si}$  & Dielectric constant of silicon                          &   11.7 F/m \\
 $\mu_{\rm{eff}}$    & Effective electron mobility                             &   $3.0\times10^{-2}$ m$^{2}$/Vs \\
 $W$                 & Width of the silicon channel                             &   $4.0\times10^{-7}$ m \\
 $L$                 & Length of the silicon channel                           &   $4.0\times10^{-7}$ m \\
 $d_y$               & Equals to $d_x$                                         &   \\
\hline
\hline
\end{tabular}
\end{center}
\begin{tablenotes}
 \item[] $^a$ Reference~\cite{Pertsev98p1988}
 \item[] $^b$ Reference~\cite{Kim05p237602}
 \item[] $^c$ Reference~\cite{taur1998fundamentals}
\end{tablenotes}
\end{table}

\section{Results and Analysis}
The ferroelectric oxide we choose is BaTiO$_3$, which possesses a relatively large spontaneous polarization ($P_s\approx$ $0.26$ C/m$^2$) at room temperature~\cite{Al-Saidi10p155304}. 

In order to simulate the energy surface, we vary the surface potential $\varphi_s$ and polarization $P_x$. 
For each $\varphi_s$, charge density $Q_z$ and polarization $P_z$ are determined uniquely by equations $\left(15\right)$ and $\left(19\right)$. 
At room temperature, the BaTiO$_3$ crystal has a tetragonal phase. The polarization orients either out of plane or in plane. 
We set the in--plane polarization direction as the $x$ direction and $P_y=0$.
Here, we should also note that we assume that the in--plane polarization has no effect on the channel. 
Therefore, it is not necessary that the source channel--drain--current flows along the $x$ direction.
Electric field $\bm{E}$ is obtained by the electrostatic restrictions in equation $\left(19\right)$.
Then energy surfaces describing Gibbs free energy $G$ with respect to $P_x$ and $P_z$ are calculated by formula $\left(1\right)$ and $\left(2\right)$. 
\begin{figure}[htbp]
\centering\includegraphics[width=15.0cm]{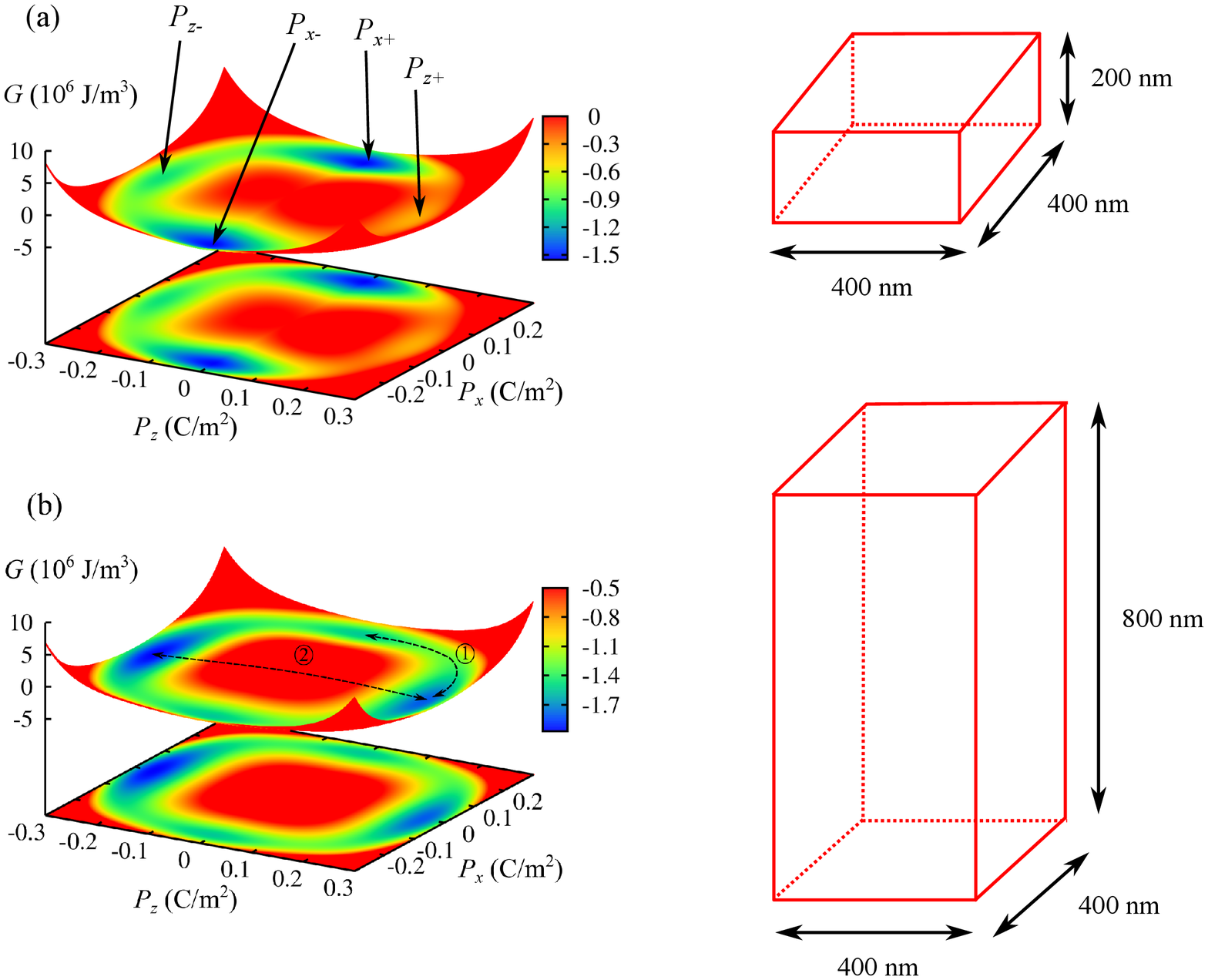} 
\caption{Energy surfaces and their two dimensional projections for MOSFET systems with ferroelectric oxide BaTiO$_3$ of different sizes.
Parameters used in the simulation are given in Ref.~\cite{Pertsev98p1988}.
$\left(a\right)$ $d_x=400$ nm and $d_z=200$ nm, polarization in plane favored. Polarizations corresponding to the four local minima are marked as $P_{x+}$, $P_{x-}$, $P_{z+}$ and $P_{z-}$;
$\left(b\right)$ $d_x=400$ nm and $d_z=800$ nm, polarization out of plane favored.
Polarization dynamics is marked with the dashed lines, path 1: polarization rotation; path 2: polarization inversion.
}\label{fig:2}
\end{figure}
In FIG. 2, we have plotted two energy surfaces of BaTiO$_3$ with different thicknesses in the $x$ and $z$ directions on a p--type silicon substrate. 

From the graphs, we can see
that for out--of--plane polarization, a negative orientation (pointing to gate electrode, with negative ends of oxide dipoles toward the channel) is more favorable when there is no applied voltage.
This is because for a p--type silicon substrate, positive screening charge is more likely to accumulate at the interface, 
leading to the polarization pointing away from the substrate/ferroelectric oxide interface. 
FIG. 3 shows the relationship of the surface potential and the interface charge density in the p--type silicon substrate. 
A positive (pointing to silicon substrate) spontaneous polarization $P_{z+}\approx0.26$ C/m$^2$ corresponds to a surface potential $\varphi_s=0.962$ eV, 
while $P_{z-}\approx-0.26$ C/m$^2$ corresponds to a surface potential $\varphi_s=-0.4346$ eV.
The depolarization fields through the ferroelectric oxide are calculated with the equation (19):
\begin{figure}[htbp]
\centering\includegraphics[width=10.0cm]{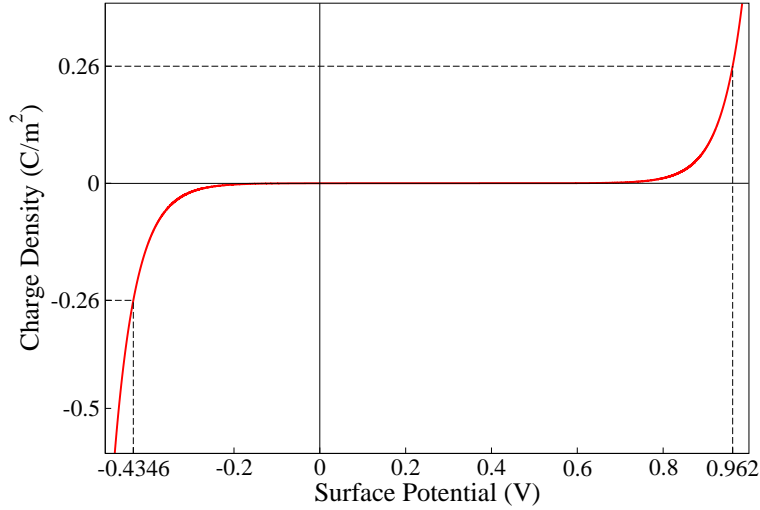} 
\caption{The relationship of surface potential and interface charge density in the doped silicon substrate. 
} \label{fig:3}
\end{figure}
\begin{equation}
E_z=\frac{1}{d_z}\left[V_g-f\left(Q_z\right)-\frac{Q_z{\lambda}_{z}}{\varepsilon_0\varepsilon_e}\right]
\end{equation}
\begin{equation}
\left|E_z\left(P_{z+}\right)\right|=\frac{1}{d_z}\left[0.962+\frac{0.26{\lambda}_{z}}{\varepsilon_0\varepsilon_e}\right]  
>\left|E_z\left(P_{z-}\right)\right|=\frac{1}{d_z}\left[0.4346+\frac{0.26{\lambda}_{z}}{\varepsilon_0\varepsilon_e}\right]
\end{equation}
The depolarization field for positive polarization is larger, and this explains why on the energy surface with no gate voltage,
a negative polarization is more favorable than a positive one.

Besides, the graphs also demonstrate the known relation that the thicker the ferroelectric oxide is in one direction, the more stable the polarization in this direction. 
As shown in equation $\left(19\right)$, if the thickness overwhelms the screening length, the potential drop in the electrodes can be neglected~\cite{Spanier06p735}. 
As a result, the electric field through the ferroelectric oxide decreases, making the polarization in this direction more favorable. 

These results also illustrate that we can modulate the global minimum by adjusting the three-dimensional size of the ferroelectric oxide. 
An energy surface we are particularly interested in possesses the global minimum for $P_x$.
When the gate voltage is applied, the local minimum corresponding to $P_{z+}$ becomes deeper and polarization rotates to the $z$ direction.
After the gate voltage is turned off, the polarization relaxes back along the $x$ direction. 
Meanwhile, the depth of the local minimum for $P_{z+}$ is close to that for $P_x$.
In such a situation, a relatively small applied gate voltage $V_g$ could induce polarization to rotate from the $x$ direction to the $z$ direction. 
The channel current strongly depends on the interface charge density,
which is approximately equal to the polarization in the $z$ direction.
\begin{equation}
I_{ds}\xrightarrow{\text{depends on}}Q_z\approx{P_z}
\end{equation}

Here, we provide guidance about how to select the optimal widths of the ferroelectric oxide, in order to make the polarization rotation likely to occur.
First, in order to make the polarization orient in the $x$ direction without gate voltage, 
the depolarization field for $P_x$ should be smaller than the one that corresponds to $P_{z-}$, 
\begin{equation}
\left|E_x\right|=\frac{1}{d_x}\left|\frac{2\times0.26{\lambda}_{x}}{\varepsilon_0\varepsilon_e}\right|        
<\left|E_z\left(P_{z-}\right)\right|=\frac{1}{d_z}\left[0.4346+\frac{0.26{\lambda}_{z}}{\varepsilon_0\varepsilon_e}\right]
\end{equation}
Typically, the dielectric constant and screening length of the noble metal electrodes are $\varepsilon_e=2$ and $\lambda=$0.4 \AA~\cite{Kim05p237602}.
For these values, we have the criterion
\begin{equation}
\frac{d_z}{d_x}<0.87
\end{equation}
In order to have a programmable device, when the applied gate voltage $V_g$ returns 0, the polarization should spontaneously return from the $P_{z+}$ position to the minimum for $P_x$ on the energy surface.
According to the Landau--Khalatnikov equation, the $P_{z+}$ position on the energy surface should be a saddle point,
\begin{equation}
\begin{aligned}
&\left.\frac{\partial^2G}{\partial{P_x}^2}<0\right|_{P_z=P_{z+}}  
&\Longrightarrow\alpha_1+\alpha_{12}P_{z+}^2+\alpha_{112}P_{z+}^4+\frac{2{\lambda}_{x}}{d_x\varepsilon_0\varepsilon_e}<0
\end{aligned}
\end{equation}
The value of $P_{z+}$ increases with thickness in the $z$ direction, 
since a thinner film means a larger depolarization field which suppresses the ferroelectricity. The $P_{z+}-d_z$ relationship is shown FIG. 4.
\begin{figure}[htbp]
\centering\includegraphics[width=10.0cm]{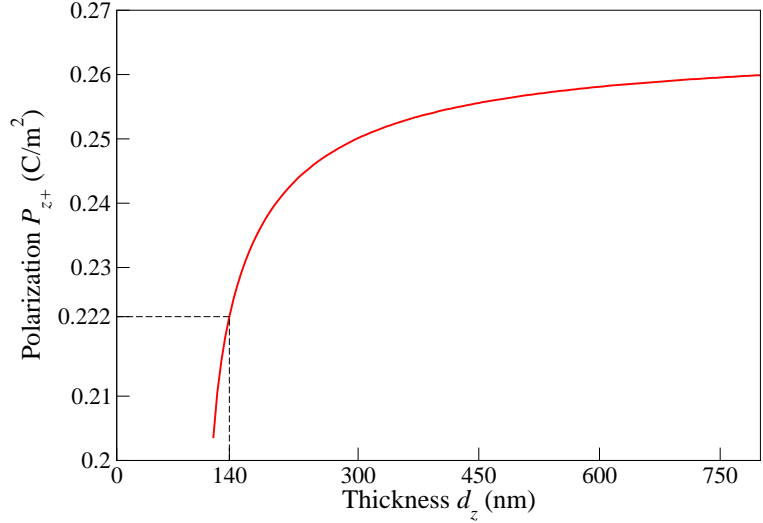} 
\caption{The $P_{z+}$ vs. $d_z$ plot. Only when the thickness in the $z$ direction $d_z$ is below 140 nm, 
$P_{z+}$ is smaller than 0.223 C/m$^2$.}
\end{figure}
\begin{equation}
\alpha_1+\alpha_{12}P_{z+}^2+\alpha_{112}P_{z+}^4<0\Rightarrow0<P_{z+}<0.223\ \rm{C/m}^2
\end{equation}
Therefore $d_z<140$ nm is a necessary condition for the polarization rotating back to the $x$ direction. 
\begin{equation}
\alpha_1+\frac{2{\lambda}_{x}}{d_x\varepsilon_0\varepsilon_e}<0\Rightarrow{d_x}>167\ \rm{nm} 
\end{equation}
According to the analysis above, in this study, the BaTiO$_3$ dimensions are selected as $d_x=400$ nm and $d_z=100$ nm.

Hysteresis loops with different values of $\gamma_0$ are calculated and shown in FIG. 5.
\begin{figure}[htbp]
\centering\includegraphics[width=10.0cm]{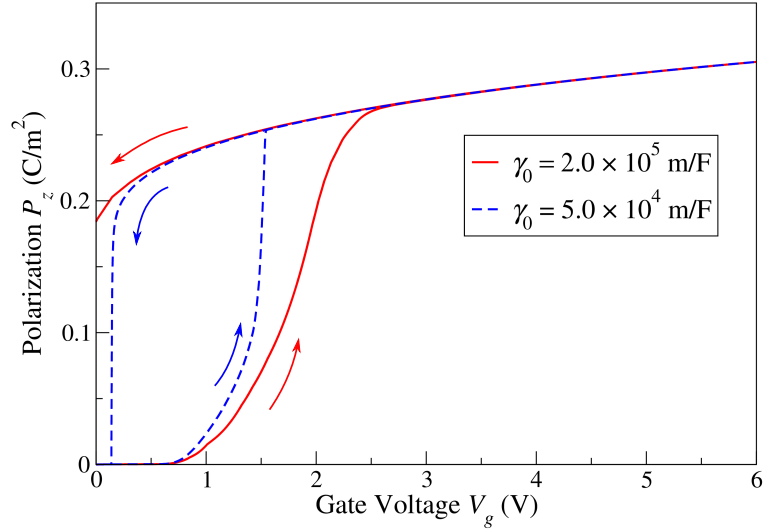} 
\caption{Hysteresis loop of the out--of--plane polarization $P_z$, with different
values of $\gamma_0$.}\label{fig:5}
\end{figure}
It demonstrates that if $\gamma_0$ is too large, the out--of--plane polarization cannot reduce to 0 and the device is not ready for the next program cycle.
From our simulation, the threshold $\gamma_0$ for out--of--plane polarization returning to 0 completely is around $1.0\times10^{5}$ m/F.
$\gamma_0$ is not only frequency dependent as shown in equation (22), but also dependent on the resistance in the circuit~\cite{Sivasubramanian03p950,Zhou05p024111},
since polarization dynamics is accompanied by screening charge transmission ~\cite{Kim05p237602,Kalinin02p3816}.
Therefore, in order to make $\gamma_0$ in the acceptable range and to have a short switching time, the resistance in the circuit should be low.
\begin{figure}[htbp]
\centering\includegraphics[width=15.0cm]{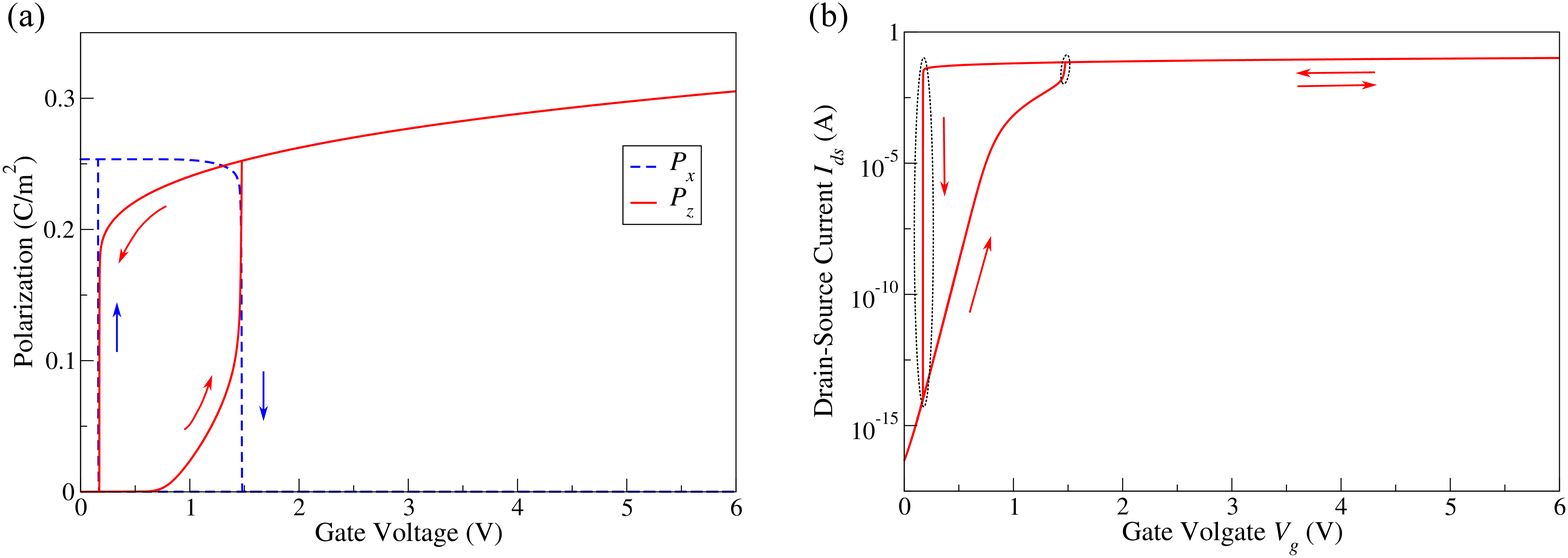} 
\caption{(a) Hysteresis loop of the in--plane polarization $P_x$ and out--of--plane polarization $P_z$,
effective polarization dynamic parameter $\gamma_0=1.0\times10^{4}$ m/F; (b) $I_{ds}-V_g$ curve of MOSFET. In the circled part, the inverse slope swing is lower than 60 mV/decade.}\label{fig:5}
\end{figure}

To evaluate the performance of MOSFET,
The drain--source current $I_{ds}$ and gate voltage $V_g$ relationship is calculated based on Pao--Sah double integral.
The simulated hysteresis loop and $I_{ds}-V_g$ curve for $\gamma_0=1.0\times10^{4}$ m/F are shown in FIG. 6 (b).

From the simulation, it can be seen that the on/off ratio of the channel current is large, which means that this device is extremely suitable for logic technology. 
This large on/off ratio results from spontaneous polarization rotation, 
because the spontaneous polarization attracts screening charge as free carriers, leading to a large on--current.
The segments in the $I_{ds}-V_g$ curve circled by dashed lines possess subthreshold swings $S$ lower than 60 mV/decade.
For the segment with $I_{ds}$ and $V_g$ increasing, $S=53$ mV/decade, and the $S$ of the decreasing segment is even lower.
This is because, as the gate voltage $V_g$ increases and exceeds the threshold voltage, the polarization rotates
and boosts free carriers in the silicon channel, 
inducing a steep increase of the channel current.
From FIG. 6, we can see that the steep change of the channel current is accompanied by polarization reorientation.

$S$ can be expressed as~\cite{Salahuddin08p405}
\begin{equation}
S=\frac{\partial{V}_g}{\partial\left(\log_{10}I_{ds}\right)}=\frac{\partial{V}_g}{\partial\varphi_s}\frac{\partial\varphi_s}{\partial\left(\log_{10}I_{ds}\right)}
\end{equation}
During the polarization reorientation period,
the polarization changes suddenly from an in--plane one corresponding to zero surface potential to a positive out--of--plane polarization, 
which maintains a large surface potential as demonstrated in FIG. 3.
The surface potential is boosted as 
\begin{equation}
\frac{\partial{V}_g}{\partial\varphi_s}<1,
\end{equation}
causing $S$ to break the 60 mV/decade limit.

Compared with polarization inversion, polarization rotation possesses many advantages for electronic device applications. 
First, as shown in FIG. 2, in the polarization rotation process, a much lower energy barrier is overcome, leading to a lower polarization rotation voltage\cite{Qi09p247603};
Secondly, the working state of the MOSFET can be modulated by a unidirectional gate voltage.
This simulation and the guidance about designing the ferroelectric oxide size can be also extended to other types of channel, such as quantum well and graphene~\cite{Liu13p053505,Ali11p1397,Baeumer14p1,Hong09p136808,Hong10p033114}.
The only part that needs to be modified according to the electric properties of new channels is the surface potential--interface charge density relationship
\begin{equation}
\varphi_s=f\left(Q_{z}\right)
\end{equation}

In the simulation, the focus is BaTiO$_3$, but this analytical model can also be applied to other 
ferroelectric oxides, such as PbTiO$_3$ and PbZr$_{1-x}$Ti$_x$O$_3$. 
PbTiO$_3$ possesses a larger energy barrier in the polarization process compared with BaTiO$_3$~\cite{Cohen92p65}. 
Therefore, a larger applied gate voltage is needed or we should use PbTiO$_3$ with smaller dimensions.
Also, a single--domain ferroelectric oxide is assumed in this model. However, the effect is not limited to a single crystal. 
When a gate voltage is applied, polarization in the $z$ direction increases in different grains and finally the polarization becomes approximately uniform. 
After the voltage is removed, the polarization relaxes back to the plane. Multiple domains may form in each grain, but the polarization distribution in plane has little effect on the channel conductance.

\section{Conclusion}
In summary, the polarization distribution in 3D and the electrical properties of the electrodes and the silicon substrate were highlighted in this LGD--theory--based model.
Our model demonstrated that polarization reorientation can modulate the drain--source current effectively.  
Besides, the choice of electrodes and the dimensions of the ferroelectric oxide are key 
factors in determining the performance of a MOSFET with depolarization fields. 
With proper selection of the thicknesses, field effect transistor with low operating voltage and fast switching can be achieved by the polarization reorientation of the ferroelectric oxide. 

\section*{ACKNOWLEDGMENTS}
Y. Q. would like to acknowledge the support of the National Science Foundation, under grant DMR-1124696. 
A. M. R. acknowledges the support of the Department of Energy, under grant DE-FG02-07ER15920.  
Both authors thank the National Energy Research Scientific Computing Center for their computational support.

\bibliography{rappecites}

\end{document}